\documentstyle[prl,multicol,aps,epsf]{revtex}
\begin{document}
\draft

\title{3D~$XY$ scaling theory of the superconducting phase 
transition}
\author{Mark Friesen and Paul Muzikar}
\address{Physics Department, Purdue University, West Lafayette,
IN 47907-1396}
\date{\today }
\maketitle
\begin{abstract}
The intermediate 3D~$XY$
scaling theory of superconductivity at zero and nonzero
magnetic fields is developed, based only upon the dimensional hypothesis
$B\sim (\mbox{Length})^{-2}$.  Universal as well as nonuniversal aspects of the 
theory are
identified, including background terms and demagnetization effects.
Two scaling regions are predicted:  an ``inner" region
(very near the zero field superconducting transition, $T_c$),
where the fields
$B$, $H$, and $H_{\rm ex}$ differ substantially, due to the presence of
diamagnetic fluctuations, and an ``outer" region (away from $T_c$),
where the fields can all be treated similarly.
The characteristic field ($H_0$) and temperature
($t_1$) scales, separating the two regimes, are
estimated.  Scaling theories of the phase transition
line, magnetization, specific heat, and conductivity are discussed.
Multicritical behavior, involving critical glass fluctuations, is 
investigated along the 
transition line, $T_m(B)$, at nonzero fields.
\end{abstract}
\begin{multicols}{2}

\section{Introduction}
The phase transition into the superconducting state in the
high-$T_c$ oxide materials has been the focus of much theoretical and
experimental research.  This transition is particularly interesting
because mean field theory does not provide an adequate
description;
for both zero and nonzero magnetic fields, fluctuations play an
important
role.  The challenge is to acquire a sound understanding of
these fluctuations in different regimes of interest.

The superconducting transition corresponds to the melting of a vortex 
solid, and occurs along the line $T_m(B)$
in the field-temperature plane.  When $B>0$, the transition can be either 
first order, as seems to be the case for pure
systems\cite{safar,liang,roulin},
or continuous, when systems contain disorder\cite{koch}.
However, we must distinguish
between the transitions at zero and nonzero fields.
For example, although $T_m(B>0)$ may be first order, the zero field
transition, $T_c$, is expected to be continuous.
The zero and finite field transitions must therefore belong to different
universality classes, as corroborated
experimentally\cite{liang,roulin}.  If the transition
$T_m(B>0)$ is also continuous, it must be associated with
a second type of critical behavior.  At low fields, it is possible to
observe both critical behaviors
simultaneously\cite{salamon,yeh,moloni,moloni5}.
The fact that the $B>0$ transition line, with its distinct critical behavior, 
connects smoothly to the zero field transition [$T_c=\lim_{B\rightarrow 
0}T_m(B)$] implies that 
($T=T_c,B=0$) is a multicritical point\cite{dfisher}.

As happens in two dimensional superconductors, the three
dimensional (3D) superconducting phase
transition occurs somewhat below the temperature where
fluctuations
of the order parameter {\it amplitude} drive the density of
fluctuating Cooper pairs to zero.  A growing body of experimental
evidence\cite{liang,roulin,salamon,yeh,moloni,moloni5,inderhees,roulin2,hubbard,cooper,kamal,anlage,ariosa,kim} 
supports the idea that the zero field transition of strongly type-II
superconductors, such as the high-$T_c$ oxides, is in the
``intermediate" 3D~$XY$
model class, which includes the $\lambda$-transition
in $^4$He.  The essential fluctuations of this class involve the order
parameter {\it phase}, but do not involve
its amplitude or the vector potential.

When vector potential fluctuations are also considered, in addition to
phase fluctuations of the order parameter, the
ultimate critical behavior of the superconducting phase transition
may belong to the ``inverted" 3D~$XY$
model class\cite{dasgupta}.
The relative stiffness of the vector potential in strongly type-II
superconductors, compared to the order parameter phase, causes this
inverted critical regime to be very
small.  In this work, we focus our attention on the intermediate
scaling region.

Critical fluctuations along the transition line $T_m(B>0)$
are thought to be of the glass type,
in the case of strong disorder\cite{koch,ffh,fisher}.
However, experiments show
that 3D~$XY$ fluctuations continue to be relevant for the scaling
when $B>0$.  These include measurements of the
specific heat\cite{roulin,salamon,inderhees,roulin2,ariosa},
magnetization\cite{liang,salamon,roulin2,hubbard,cooper},
penetration depth\cite{kamal}, and
conductivity\cite{salamon,yeh,moloni,moloni5,ariosa,kim}.
While 3D~$XY$ and glass fluctuations may
coexist at low fields near the superconducting transition, they correspond 
to two distinct universality classes, with distinct
exponents and scaling functions.  This multicritical coexistence of
superconducting fluctuations has been studied
experimentally\cite{salamon,yeh,moloni,moloni5}.

In this paper we consider intermediate
3D~$XY$ fluctuations, for low magnetic fields, $B\geq 0$\cite{notezlatko}.
For simplicity we consider only the case of a continuous transition 
$T_m(B)$ at nonzero fields; details of the case of first order melting
are presented elsewhere\cite{twophase}.
Some other theoretical treatments of the 3D~$XY$ model of superconductivity 
include Refs.~\cite{ariosa,ffh,lobb,zlatko,schneider,lawrie}, while
some relevant numerical simulations are given in Ref.~\cite{numerics}.
Although exact solutions to this problem are not yet in sight, a
fruitful advance can be made using the scaling approach.  In this
method, a scaling form for the free energy is hypothesized
by means of a dimensional analysis.  Using this ansatz, the
theory is developed in a very general form,
relying as little as possible on particular models for the
superconductor.  In addition to the free energy, we discuss the
magnetization, specific heat, and conductivity.
One important aspect of the present work is to clearly specify
the nonuniversal parameters which enter the theory.  These are of
three types.  One type comes from the smooth background terms.  The 
second type is associated with demagnetization effects, and is related to 
sample geometry.  The third type involves material-dependent constants 
which enter the scaling term of the free energy.

We also carefully distinguish between the different fields:  the spatially 
averaged
magnetic field $B$, its conjugate field $H$, and the external field
$H_{\rm ex}$.  The usual assumption that $B\simeq H_{\rm ex}$, 
appropriate
for a disk geometry, is reconsidered.  The distinct thermodynamic roles 
of
$B$ and $H$, defined through the relation $H=4\pi \partial f/\partial B$
($f$ is the appropriate free energy density), suggest that the two
quantities should scale differently; this reflects the emergence of
diamagnetic fluctuations.  The distinction between the different
fields is most apparent in a small ``inner" scaling region near
the zero field transition.  In a realistic physical scenario,
$H_{\rm ex}$ is the externally controlled variable; $B$ and $H$ then 
both
acquire fluctuation contributions.  Since estimates of the inverted
$XY$ scaling regime place it in the vicinity of the inner scaling
region, it is therefore crucial to elucidate the differences between
the different fields,
in order to unravel the different 
critical phenomena.

The plan of the paper is as follows.
Section~II discusses the scaling form of the free energy,
Eq.~(\ref{eq:ftilde}).  Section~III derives the magnetic equation of state,
Eq.~(\ref{eq:H}).  The crossover field, $H_0$, between the inner and
outer scaling regions, is identified in Eq.~(\ref{eq:H0}).  Section~IV
derives the form of the
superconducting phase transition line in the
$B$-$T$, $H$-$T$, and $H_{\rm ex}$-$T$
planes, given by Eqs.~(\ref{eq:BT}), (\ref{eq:HTp}), and
(\ref{eq:HexTp}), respectively.  The $H$-$T$ phase diagram is shown 
in Fig.~1.  Section~V is a brief interlude, which shows how the
Abrikosov theory of the superconducting transition in a field is a
special case of
the more general scaling theory.  The magnetization is discussed in
Section~VI, where we derive the relations between $M$ and $H$
[Eq.~(\ref{eq:MofH})] or $M$ and
$H_{\rm ex}$ [Eq.~(\ref{eq:MofHex})], for $T=T_c$ or $T_m$.
In Section~VII we obtain the specific heat, Eq.~(\ref{eq:C}).  
Section~VIII postulates the
dynamic scaling theory associated with the ohmic
conductivity, Eq.~(\ref{eq:sigma}).  In
Section~IX we conclude, giving estimates for the size of the inner
and inverted scaling regions.  

{\it Notation:}  In this paper, only the quantities with tildes involve 
factors related to sample geometry.

\section{Scaling Theory}
We now discuss the basic thermodynamics of the scaling 
approach.  At zero field, we adopt the usual scaling hypothesis, which 
states that any observed
singular behavior involves the divergence of the correlation length, 
in 
terms of the relative temperature 
$t\equiv (T-T_c)/T_c$.  The zero field correlation length 
$\xi (T)$ and specific heat $C(T)$ are then described by
power laws, with the exponents $\nu_{xy}$ 
and $\alpha_{xy}$, respectively\cite{goldenfeld}.

To extend scaling to finite fields, we must know how $B$ scales.  We 
therefore introduce a definite assumption about the physics of a 
superconductor.  A defining characteristic of a superconductor is its 
broken $U(1)$ or gauge symmetry, which is reflected in the 
symmetry of the
superconducting order parameter.  Gauge-invariance then implies the
following identification for the gradient operator:
$\bbox{\nabla} \rightarrow {\bf D}=\bbox{\nabla}+2ie{\bf A}/\hbar c$.  
The basic scaling argument, which amounts to a
dimensional analysis, states that the two terms appearing in $\bf D$ must 
have the same scaling dimension:
$(\mbox{Length})^{-1}$.  Similar dimensional arguments were 
proposed in 
Ref.~\cite{ffh}, and later confirmed in Ref.~\cite{lawrie}. 
The dimensionality of the magnetic field is then expressed as
\begin{equation}
B=|\bbox{\nabla} \times {\bf A}|\sim (\mbox{Length})^{-2}. \label{eq:bscale}
\end{equation}

The most general scaling 
hypothesis for the free energy 
density becomes\cite{note1}
\begin{equation}
f(B,T)=f_b(B,T)+f_k|t|^{2-\alpha_{xy}}
\phi_{\pm} \left( \frac{B|t|^{-2\nu_{xy}}}{H_k} \right) 
.\label{eq:ftilde}
\end{equation}
The function $f_b(B,T)$ represents the 
smooth background, while the second term in Eq.~(\ref{eq:ftilde}) 
encapsulates all the effects of 3D~$XY$ critical fluctuations.  The 
scaling theory is universal in the sense that neither the exponents,
$\nu_{xy}$ and $\alpha_{xy}$, nor
the functions $\phi_+$ ($\phi_-$), corresponding to $t>0$ ($t<0$), 
contain any sample dependence; they are the same for all 
superconductors which exhibit 3D~$XY$ scaling.  The sample 
dependence rests only in the background term $f_b(B,T)$, the transition 
temperature $T_c$, and in the 
parameters $f_k$ and $H_k$ in the fluctuation term.  Here, $f_k$ has 
units of free energy 
density, making 
$\phi_\pm (x)$ dimensionless, while $H_k$ has field units, making 
the scaling variable,
$x=B|t|^{-2\nu_{xy}}/H_k$, dimensionless.  

We now discuss several points concerning the free energy, 
Eq.~(\ref{eq:ftilde}):

(1)  The arguments leading to Eq.~(\ref{eq:bscale}) do 
not admit anomalous scaling dimensions, since
the magnetic field does not fluctuate in the
intermediate $XY$ model class. On the other hand, 
fluctuations of the vector potential become important
in the inverted~$XY$ critical region, and may lead to the 
appearance of an anomalous dimension:
$B\sim (\mbox{Length})^{-2 +\vartheta }$.  
Mounting experimental evidence seems to 
be consistent with the intermediate scaling presciption of $\vartheta =0$.  
In the 
work which follows, 
the magnetic field is always treated within the intermediate scaling
hypothesis.  In particular, we emphasize that the inner scaling region, 
discussed below, also emerges from Eq.~(\ref{eq:ftilde}), 
and is not related to inverted 3D~$XY$ behavior.

(2)  The region of the $B$-$T$ plane for which 
Eq.~(\ref{eq:ftilde}) provides the correct scaling description must either be 
determined experimentally, or by a more detailed theory.  
Empirically, the range of validity of Eq.~(\ref{eq:ftilde}) appears to 
extend far beyond the 
characteristic 3D~$XY$ temperature and field scales, $t_1$ and $H_0$ 
(discussed below), which are very 
small in many materials.

(3)  Near $T=T_c$, the background term in $f$ may be written 
approximately as follows:
\begin{equation}
f_b(B,T)\simeq f_0(T)+f_2(T)B^2 ,\label{eq:fbk}
\end{equation}
where $f_0(T)$ and $f_2(T)$ are smooth 
functions, with no singular 
behavior.  We shall see that $f_2(T)$ is 
related to the 
background magnetic susceptibility.

(4)  The sample dependent quantities defined above, $T_c$, $f_k$, 
$H_k$, and $f_b(B,T)$, must all reflect the anistropy of the 
superconductor.  The main effect of anisotropy on Eq.~(\ref{eq:ftilde}) 
is that $H_k^{-1}$ becomes a tensor quantity.
In other words, anisotropy causes $H_k$ to depend on the 
direction of $\bf B$.  To avoid such complications here, we may 
simplify the analysis by choosing our geometry carefully:
we consider the simple, but experimentally relevant 
case that anisotropy, if present, is aligned with the principal axes 
of an assumed ellipsoidal sample.  In addition, we require that any 
external magnetic field should be applied along a principal axis of 
the sample.  In other geometries, the field $\bf B$ becomes 
nonuniform and/or misaligned with the external field 
${\bf H}_{\rm ex}$.

(5)  It is possible to relate the parameter $f_k$ to
another nonuniversal parameter $\xi_0$, which appears in the zero field
critical correlation length $\xi (T)=\xi_0 |t|^{-\nu_{xy}}$, by means of
two-scale-factor universality\cite{schneider,privman}.  The result can be 
written as $f_k=k_BT_c/\xi_0^3$, where for simplicity, we have 
absorbed
a universal proportionality constant into the definitions of $f_k$ and
$\xi_0$.  Without loss of generality, we may also introduce anisotropy
into this relation as $f_k=\gamma k_BT_c/\xi_{ab0}^3$, 
where $\gamma =\xi_{ab0}/\xi_{c0}$ represents the ratio of the zero
field 3D~$XY$ critical correlation lengths along different axes.

Several theoretical analyses of superconductors relate the
renormalized anisotropy parameter, $\gamma$, to the anisotropy
appearing in the bare 
Hamiltonian, by showing that the anisotropic problem may be treated
as an isotropic one\cite{ariosa,blatter}.  In this paper, 
we also assume that the isotropic and anisotropic problems should involve
the same, universal scaling functions.  However, we work strictly
in terms of the measurable (renormalized) anisotropy factor.
Note that unless $\xi_{ab0}$ can be determined by independent means, 
it is not 
possible to extract $\gamma$ directly from 
Eq.~(\ref{eq:ftilde}), 
without making further assumptions.  (These are described in the 
following 
point.)  We therefore retain the more general ($f_k,H_k$) notation here,
noting that anisotropy is naturally absorbed into these parameters.

(6)  The scaling ansatz of Eq.~(\ref{eq:ftilde}) is very general,
since it arises from purely dimensional arguments.  This form involves
exactly two nonuniversal parameters\cite{privman}, $f_k$ and $H_k$, 
in addition to the temperature scale, $T_c$, and the background
term, $f_b(B,T)$.  It is not possible to reduce this number of sample
dependent parameters without further information.  Furthermore, we 
emphasize that the 3D~$XY$ model has not been solved exactly, and
therefore cannot provide such information.  However, it has been 
suggested that a relation exists between $f_k$ and 
$H_k$\cite{ffh,schneider}, thus reducing the number of  nonuniversal 
parameters by one.  

The heuristic argument states that
a characteristic field scale, $B_{\rm ch}(T)$, appears in the argument 
of the scaling functions $\phi_\pm (x)$, in the form 
$x=B/B_{\rm ch}(T)$.  This field scale should be given
precisely by\cite{ffh,schneider} $B_{\rm ch}(T)=\Phi_0/\xi^2(T)$, 
where $\xi (T)$ is again the zero field correlation length.
It follows that 
$H_k=\Phi_0/\xi_0^2$, from which we obtain the desired relation: 
$f_k/k_BT_c=(H_k/\Phi_0)^{3/2}$.
For anisotropic  superconductors, the relation becomes
$f_k/\gamma k_BT_c=(H_k/\Phi_0)^{3/2}$, 
when $\bf B||\hat{c}$, thereby providing a method for determining 
the anisotropy, $\gamma$, when the parameters $f_k$ and $H_k$ are 
determined experimentally.
 
The proposed relation between $f_k$ and $H_k$ can be 
tested through
a scaling analysis of the fluctuation mangetization, as described in
Section~VI.  We point out that
the relation between certainly does not hold for the
case of the mean field Abrikosov theory, described in Section~V.  
However, for critical fluctuations, the question is still open.
For generality here, we continue to treat $f_k$ and $H_k$ as 
independent parameters.

\section{Equation of State}
To procede with the analysis of Eq.~(\ref{eq:ftilde}), the hyperscaling 
relation may be used:  $2-\alpha_{xy}=3\nu_{xy}$ (for the 3D case).
Eqs.~(\ref{eq:ftilde}) and~(\ref{eq:fbk}) can be rewritten as 
\begin{equation}
f(B,T)=f_0(T)+f_2(T)B^2 +
f_k|t|^{3\nu_{xy}}\phi_\pm 
\left( \frac{B|t|^{-2\nu_{xy}}}{H_k}\right) .\label{eq:f}
\end{equation}
The magnetic equation of state is derived from the identity
$H=4\pi \, \partial f /\partial B$:
\begin{equation}
H=\Omega_T B+
\left( \frac{4\pi f_k}{H_k} \right) |t|^{\nu_{xy}} \phi'_{\pm} 
\left( \frac{B|t|^{-2\nu_{xy}}}{H_k} \right)  , 
\label{eq:H}
\end{equation}
where $\phi'_{\pm}(x)=\partial \phi_{\pm}/\partial x$.  
Eq.~(\ref{eq:H}) relates $B$ and $H$ for a 
given temperature.  The normal and  fluctuation contributions 
appear 
in the first and second terms, respectively.  
$\Omega_T\equiv 8\pi f_2(T)$ is the inverse permeability of 
the background; experimentally, it is found that 
$\Omega_T\simeq 1$.  

It is desirable to invert Eq.~(\ref{eq:H}) to obtain $B(H,T)$.  In general, 
this is not possible, because $\phi_{\pm}(x)$ are not yet known 
theoretically.  However, progress can be made in certain cases.
When $B>0$ and $T=T_c$, the fluctuation part of Eq.~(\ref{eq:H}) 
must be smooth and independent of $t$.  This leads to the following 
asymptotic behavior:
\begin{equation}
\lim_{x\rightarrow \infty} 4\pi \phi'_\pm (x)
=(b_0x)^{1/2} ,
\label{eq:phiprime}
\end{equation}
where, $b_0$ is a dimensionless, universal constant of
the 3D~$XY$ theory.
Thus when $T=T_c$, we have
\begin{equation}
\Omega_cB=H+2H_0-2H_0^{1/2}\sqrt{H_0+H}.\label{eq:Hlimit}
\end{equation}
Here we have simplified the notation using 
$\Omega_c\equiv \Omega_{T_c}$.

In Eq.~(\ref{eq:Hlimit}) we have defined a characteristic field 
\begin{equation}
H_0\equiv b_0f_k^2/4\Omega_cH_k^3, \label{eq:H0}
\end{equation}
which represents the crossover between the inner and outer scaling 
regions, for 
$T=T_c$.  This crossover field is 
very small for most superconductors, due to the small size of 
diamagnetic fluctuations above the transition.  In the outer scaling 
region, $H\gg H_0$, we see that the usual approximation 
$H\simeq \Omega_cB$ is appropriate.  However, in the inner 
region, $H\ll H_0$, diamagnetic fluctuations dominate the equation of 
state, 
leading to the behavior $B\propto H^{1/2}$, for $T=T_c$.  If there
exists an exact
relation between $f_k$ and $H_k$, as described in Sec.~II, then 
Eq.~(\ref{eq:H0}) should reduce to $H_0\propto \gamma^2T_c^2$,
demonstrating the strong anisotropy dependence of the inner scaling region 
becomes explicit.

\section{Phase Transition Line}
The free energy (\ref{eq:f}) must contain information concerning the 
phase transition into the superconducting state.  Of crucial importance 
is the fact that the superconducting transition line $T_m(B)$ 
terminates at $T_c$ on the $B=0$ axis.  The special point 
($T=T_c,B=0$) is then multicritical\cite{riedel}.  For the case of a 
continuous transition considered in this paper,
the transition always occurs at a particular, universal value 
of the scaling variable:  $x=x_m$.  
The transition line is then given by
\begin{equation}
B(T) =(x_mH_k)|t|^{2\nu_{xy}} .\label{eq:BT}
\end{equation}

Combining Eqs.~(\ref{eq:H}) and (\ref{eq:BT}), we obtain the 
transition line in 
the $H$-$T$ plane:
\begin{equation}
H(T)=\Omega_Tx_mH_k|t|^{2\nu_{xy}} +
\frac{4\pi f_k}{H_k} \phi'_- (x_m) |t|^{\nu_{xy}} .\label{eq:HTp0}
\end{equation}
Eq.~(\ref{eq:HTp0}) has two terms, one going as $|t|^{\nu_{xy}}$, 
and the other 
going as $|t|^{2\nu_{xy}}$.  Two different types of 
nonuniversal parameters enter 
the $|t|^{2\nu_{xy}}$ term, including the background quantity 
$\Omega_T$, which contains a possible temperature dependence.  For 
simplicity, we will assume that $\Omega_T$ 
is nearly constant over the temperature range of interest:  
$\Omega_T\simeq \Omega_{T_c}\equiv \Omega_c$.

Eq.~(\ref{eq:HTp0}) then becomes
\begin{equation}
\frac{H(T)}{\Omega_cx_mH_k} =|t|^{2\nu_{xy}}+
(t_1|t|)^{\nu_{xy}} .\label{eq:HTp}
\end{equation}
We have introduced the relative temperature scale
\begin{equation}
t_1\equiv (f_kb_1/\Omega_cx_mH_k^2)^{1/\nu_{xy}}, \label{eq:t1}
\end{equation}
and the universal number
$b_1\equiv 4\pi \phi'_-(x_m)$.  The expression $t_1T_c$ represents the 
size of the inner scaling region along the transition line $T_m(H)$;
it is analagous to $H_0$ on the $H$~axis.  The 
$H$-$T$ phase diagram is sketched in Fig.~1.

\vspace*{-.5truein}

\begin{figure}
\epsfxsize=3.1truein
\vbox{\hskip 0truein
\epsffile{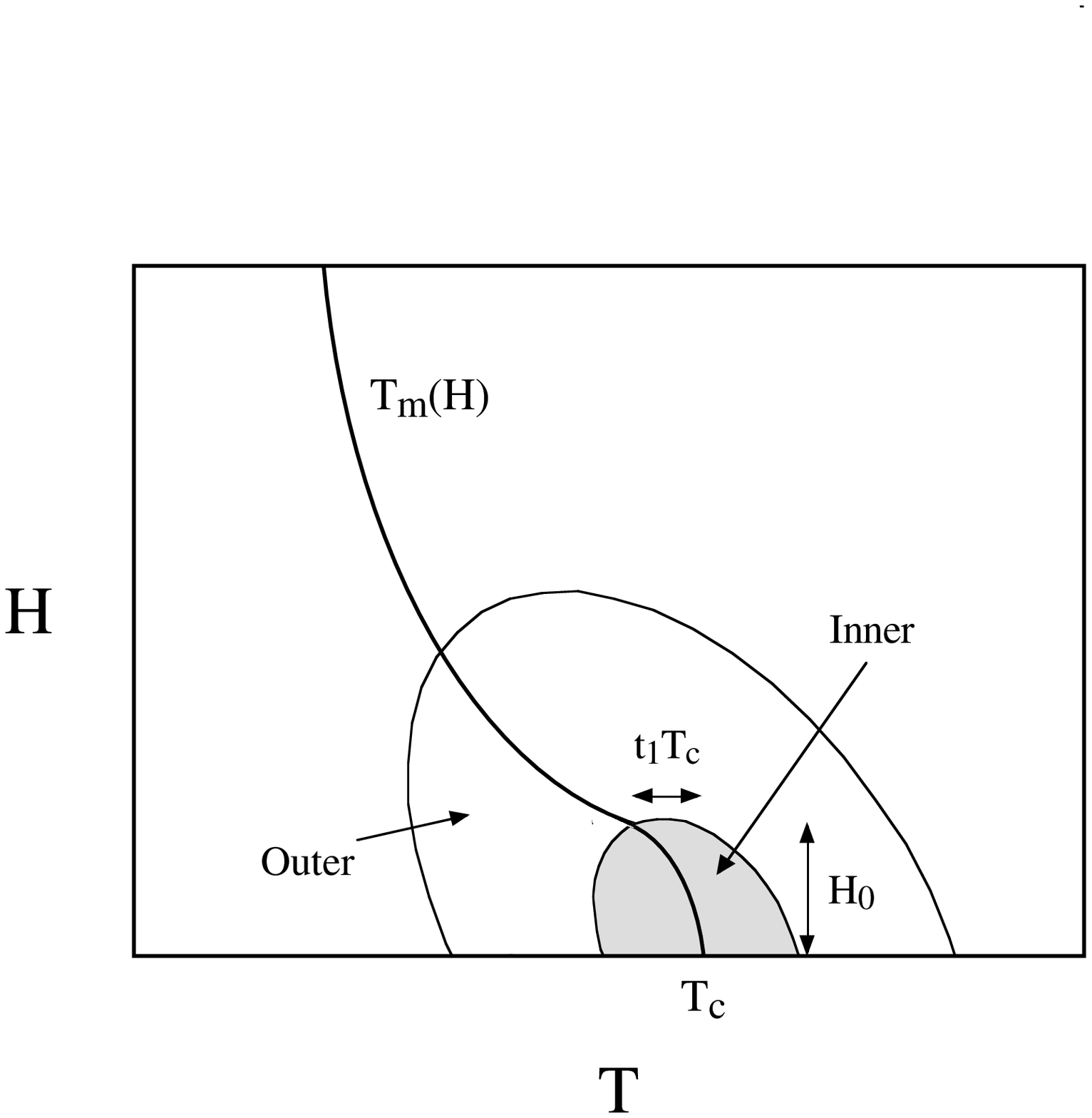}}
\end{figure}
\vspace{-.4truein}
\centerline{\begin{minipage}{3in}
\small \noindent FIG.~1.
Phase diagram in the $H$-$T$ plane, showing the phase transition
line $T_m(H)$ and the different scaling regions.  In the shaded, ``inner"
scaling region, the usual
approximation $B\simeq H$ breaks down.  The dimensions of the inner
region are of
order $H_0 \times t_1T_c$, where $H_0$ is defined in
Eq.~(\ref{eq:H0}) and $t_1$ in Eq.~(\ref{eq:t1}).  In the unshaded,
``outer" region, the approximation $B\simeq H$ is
accurate.  Relative sizes of the two scaling regions are not drawn to
scale; in many cases, the inner region is tiny compared to the outer
region.
\end{minipage}}\vspace{.3in}

It is also of interest to study the phase transition line in the 
$H_{\rm ex}$-$T$ plane, because of its experimental importance.  We 
assume that $H_{\rm ex}$ is applied 
parallel to a principal axis of our sample, which is an 
ellipsoid\cite{note4}.  A simple equation then relates the different 
fields\cite{landau}:
\begin{equation}
H_{\rm ex}=nB+(1-n)H ,\label{eq:demag}
\end{equation}
where $n$ is the demagnetizing coefficient, satisfying 
$0\leq n\leq 1$.  ($n\simeq 0.8-0.9$ for typical high-$T_c$ single 
crystals, 
while $n\gtrsim 0.99$ for thin films.)  Eqs.~(\ref{eq:BT}), 
(\ref{eq:HTp}), and (\ref{eq:demag}) then give the transition 
line:
\begin{equation}
\frac{H_{\rm ex}(T)}{\tilde{\Omega}_c x_mH_k} 
=|t|^{2\nu_{xy}} +(\tilde{t}_1|t|)^{\nu_{xy}} ,\label{eq:HexTp}
\end{equation}
where $(1-\tilde{\Omega}_T)\equiv (1-n)(1-\Omega_T)$, and 
$\tilde{t}_1^{\, \nu_{xy}}\equiv t_1^{\nu_{xy}}(1-n)\Omega_c/
\tilde{\Omega}_c$.  Note again that quantities with tildes differ from 
those without tildes only by geometric factors.

The limiting behaviors of Eqs.~(\ref{eq:HTp}) and (\ref{eq:HexTp}) 
are clear; we show only the result for $H_{\rm ex}$:
\begin{equation}
H_{\rm ex} \simeq \left\{ \begin{array}{lcl} \,
(\tilde{\Omega}_c x_mH_k\tilde{t}_1^{\, \nu_{xy}}) 
|t|^{\nu_{xy}}
& \quad & |t| \ll \tilde{t}_1 , \\
\, (\tilde{\Omega}_c x_mH_k) |t|^{2\nu_{xy}}
& \quad & |t| \gg \tilde{t}_1 .
\end{array} \right. \label{eq:Hexas}
\end{equation}
The large $|t|$ limit corresponds to the usual (outer region)
description of the 
phase boundary\cite{ffh}.  New behavior is observed in the small inner 
region, $|t| \ll t_1$. 

\section{Mean Field Theory}
As noted in the introduction, the superconducting transition in high-$T_c$
materials should not be amenable to mean field description.  However in
low-$T_c$ superconductors, the mean field description is often found to
be accurate.
It is therefore interesting to note that the Abrikosov theory\cite{degennes} of 
the $B> 0$ 
mean field transition is consistent with the scaling theory presented 
so far.
We now pause briefly to discuss this point.  

Beginning with Eq.~(\ref{eq:ftilde}), progress cannot be 
made using hyperscaling, since this relation does not hold for the 
mean 
field transition.  However, mean field 
exponents may be used, 
since they are known exactly:  $\nu_{\rm mf}=1/2$ and 
$\alpha_{\rm mf}=0$.  This gives
\begin{equation}
f_{\rm mf}=f_0(T)+\frac{B^2}{8\pi} +f_k|t_0|^2\phi_{\rm mf \pm}
\left( \frac{B|t_0|^{-1}}{H_k} \right) ,\label{eq:fmf2}
\end{equation}
where $t_0\equiv (T-T_{c0})/T_{c0}$, and $T_{c0}$ refers to the 
temperature where the upper critical field vanishes:  
$B_{c2}(T=T_{c0})=0$.
For simplicity, we have assumed that the normal state background 
has no 
magnetic effects ($\Omega_T=1$), as is usual in the Abrikosov 
theory.  The equation of state then becomes
\begin{equation}
H=B+\frac{4\pi f_k}{H_k} |t_0|\phi'_{\rm mf\pm}
\left( \frac{B|t_0|^{-1}}{H_k} \right) .\label{eq:Hmf2}
\end{equation}
The mean field phase transition, $B_{c2}(T)$, occurs when the 
argument of $\phi_{\rm mf-}(x)$ has the value $x_m$, leading to the 
following transition line in the $B$-$T$ plane:
\begin{equation}
B=(x_mH_k)|t_0| .\label{eq:Bc2}
\end{equation}
Using Eq.~(\ref{eq:Hmf2}), the transition line in the $H$-$T$ plane is 
given by 
\begin{equation}
H=\left[ x_mH_k+\frac{4\pi f_k}{H_k} 
\phi'_{\rm mf-}(x_m) \right] \frac{T_{c0}-T}{T_{c0}} 
.\label{eq:Hc2}
\end{equation}
The linearity of the transition line in $T_{c0}-T$ is consistent with
the Abrikosov theory.  We 
emphasize that 
this prediction, Eq.~(\ref{eq:Hc2}), has been obtained with no 
explicit 
knowledge of the Abrikosov ($B>0$) solution.  Instead, it is a 
general consequence of the scaling theory. 

The scaling functions $\phi_{\rm mf \pm}(x)$ can now be explicitly 
computed.  Recall the Abrikosov solution for the free 
energy,
$f_{\rm mf}(B,T)$, near the  upper critical field 
$H_{c2}(T)$\cite{degennes}:
\begin{equation}
f_{\rm mf} -f_0=\left\{ \begin{array}{ccl}
\frac{B^2}{8\pi}-\frac{1}{8\pi}
\frac{(H_{c2}-B)^2}{1+(2\kappa^2-1)\beta_A}
& \quad & B\lesssim H_{c2}(T) \\
\frac{B^2}{8\pi} & \quad & B>H_{c2}(T) 
\end{array} \right.  
,\label{eq:fmf3}
\end{equation}
where
\begin{equation}
H_{c2}(T)\equiv \frac{\Phi_0}{2\pi \xi_{\rm BCS}^2}
\frac{T_{c0}-T}{T_{c0}} .\label{eq:Hc23}
\end{equation}
Here, $\kappa$ is the Ginzburg parameter, $\beta_A\simeq 1.16$ 
for the triangular vortex lattice, and 
$\xi_{\rm BCS}\sim \hbar v_F/k_BT_{c0}$ is 
the temperature independent coherence length.  

The correspondence between Eqs.~(\ref{eq:fmf2}) and (\ref{eq:fmf3}) 
becomes transparent by 
making the following identifications.  The nonuniversal parameters can
be taken as $H_k=\Phi_0/2\pi \xi_{\rm BCS}^2$ and
$f_k=H_k^2[1+(2\kappa^2+1)\beta_A]^{-1}/8\pi$,
where we have adopted the following normalization:  $\phi_{\rm mf -}(0)=-1$. 
The quantities $f_k$ and $H_k$
contain all the sample dependence of the scaling description.  
Comparison with 
Eq.~(\ref{eq:fmf2}) now gives
\begin{eqnarray}
\phi_{\rm mf -}(x) & = & \left\{ \begin{array}{ccl}
-(1-x)^2 & \quad & x\lesssim 1 \\ 
0 & \quad & x>1 \\ 
\end{array} \right.  ,\label{eq:phimfm} \\ 
\phi_{\rm mf +}(x) & = & 0 . \label{eq:phimfp}
\end{eqnarray}
Within the mean field approach, we can evaluate the universal 
quantities $b_1$ and $x_m$ introduced earlier.  We find that 
$b_1=0$, reflecting the lack of fluctuations in this case.  (At 
temperatures above the transition, we simply have $B=H$.)  
Additionally, we find $x_m=1$.

Finally, we point out that mean-field theory provides an example of a 
case where the nonuniversal parameters, $f_k$ and $H_k$, cannot 
be related in a simple way, except by introducing an additional, 
nonuniversal parameter $\kappa$.  However, $H_k$ does take the form
suggested by heuristic arguments in Sec.~II.
Note that the relation found here, 
$f_k\propto H_k^2$, differs from the one described in Sec.~II, due to 
the breakdown of hyperscaling in mean field theory.

\section{Magnetization}
In the remainder of this paper we consider only 3D~$XY$ 
critical fluctuations.  As usual, the magnetization is defined by
$M=(B-H)/4\pi$.
Using Eq.~(\ref{eq:H}), we obtain
\begin{equation}
M(B,T)=\frac{1-\Omega_T}{4\pi}B-
\frac{f_k}{H_k^{3/2}} B^{1/2}{\cal M}_{\pm}\left( 
\frac{B|t|^{-2\nu_{xy}}}{H_k} \right) 
,\label{eq:Mscale}
\end{equation}
where ${\cal M}_{\pm}(x)= x^{-1/2}\phi'_{\pm}(x)$ are universal scaling 
functions.
The first term in Eq.~(\ref{eq:Mscale}) represents the normal state 
background, and vanishes for $\Omega_T=1$.  The second 
term is the critical fluctuation contribution. 

As discussed in Sec.~II, certain physical arguments may lead to a 
reduced number of nonuniversal parameters, by providing a relation
$f_kH_k^{-3/2}\propto \gamma T_c$ ($\gamma =1$ for the isotropic 
case), where the proportionality constant is not sample dependent.  
Eq.~(\ref{eq:Mscale}) offers a convenient experimental test 
of this 
prediction, since the expression
$f_kH_k^{-3/2}$, can be directly infered
from the scaling.  To test the prediction, the quantities $T_c$ and 
$\gamma$ should be 
determined by independent methods\cite{note5}.

Deriving equations for $M(H,T)$ or $M(H_{\rm ex},T)$ is not 
straightforward, because of the difficulty in inverting
Eq.~(\ref{eq:H}) to obtain $B(H,T)$.  We therefore limit our
derivation to the formulae relating $M$ and $H$, or $M$ and 
$H_{\rm ex}$, at the special temperatures $T_c$ and $T_m$.
Using the appropriate equations of state, we find the following 
results at $T=T_c$:
\begin{eqnarray}
4\pi \Omega_cM & = & (1-\Omega_c)H+2H_0
\nonumber \\ & & \mbox{} \,
-2H_0^{1/2}\sqrt{H_0+H}, \label{eq:MofH} \\
4\pi (1-n) \tilde{\Omega}_cM & = &
(1-\tilde{\Omega}_c)H_{\rm ex}+2\tilde{H}_0
\nonumber \\ & & \mbox{} \,
-2\tilde{H}_0^{1/2}\sqrt{\tilde{H}_0+H_{\rm ex}}
,\label{eq:MofHex}
\end{eqnarray}
where 
$\tilde{H}_0\equiv H_0(1-n)^2\Omega_c/\tilde{\Omega}_c$.  
The first term on the right hand side of both equations gives the normal 
background contribution, while the 
remaining terms represent the fluctuations.  The asymptotic 
behavior of the fluctuation part of 
the magnetization, $M_{\rm fl}$, can be found.  We 
show the results for $H_{\rm ex}$:
\begin{equation}
M_{\rm fl}\simeq \left\{ \begin{array}{ccl}
-\frac{H_{\rm ex}}{4\pi (1-n)\tilde{\Omega}_c} & \quad &
H_{\rm ex}\ll \tilde{H}_0 \\ 
-\frac{\tilde{H}_0^{1/2}H_{\rm ex}^{1/2}}
{2\pi (1-n)\tilde{\Omega}_c}  &\quad &
H_{\rm ex}\gg \tilde{H}_0 
\end{array} \right. .\label{eq:Mfl}
\end{equation}
In the low field limit, $M_{\rm fl}$ becomes 
asymptotically linear in $H_{ex}$, 
like the background.  The fluctuation 
magnetization then dominates over the background by 
a factor proportional to $(1-n)^{-1}(1-\tilde{\Omega}_c)^{-1}$, which 
can be quite large for typical (flat) samples.  
Thus, in the low field ($H_{\rm ex}\ll \tilde{H}_0$), inner scaling 
region, the 
problem of background subtraction, which otherwise troubles 
experimental analyses, is ameliorated.

When $T=T_m$, we may still use Eqs.~(\ref{eq:MofH}) 
and~(\ref{eq:MofHex}) by making the following replacements:
\begin{equation}
H_0\rightarrow \left( \frac{b_1^2}{x_mb_0} \right) H_0
\quad \mbox{and} \quad
\tilde{H}_0\rightarrow \left( \frac{b_1^2}{x_mb_0} \right) 
\tilde{H}_0
.\label{eq:replace}
\end{equation}

\section{Specific Heat}
The free energy (\ref{eq:f}) may be used to compute the specific heat 
at constant $B$:
\begin{eqnarray}
\frac{C(B,T)}{T} & = & -f''_0(T)-
f''_2(T) B^2
\nonumber \\ & & \mbox{} \,
+f_kT_c^{-2}|t|^{-\alpha_{xy}}
\psi_\pm \left( \frac{B|t|^{-2\nu_{xy}}}{H_k} \right) .\label{eq:C}
\end{eqnarray}The dimensionless functions $\psi_\pm (x)$ 
depend on $\phi_\pm (x)$, and their first and second derivatives.  
The first two terms in Eq.~(\ref{eq:C}) represent the background, 
while the last term is 
due to superconducting fluctuations.  

We note the following points:

(1)  Experimentally, it is difficult to isolate the fluctuation 
contributions in Eq.~(\ref{eq:C}), due to
(i) the smallness of fluctuations compared to the background,
(ii) the weakness of the specific heat singularity\cite{zinn} 
($\alpha_{xy} \simeq -0.01$), 
(iii) rounding effects, which are often observed 
in experiments.

(2)  It is possible to simplify Eq.~(\ref{eq:C})
for the high-$T_c$ materials, by using the empirical fact that 
$f''_2(T)\simeq 0$.  The following scaling quantity may then 
be considered\cite{salamon}:
\begin{equation}
\Delta C(B,T)\equiv C(B,T)-C(0,T) ,\label{eq:deltaC}
\end{equation}
which contains no background dependence.  However, if 
the zero field specific heat cusp becomes rounded near $T_c$, as is 
often the case for real samples, then 
the imperfect scaling, associated with the rounding, is transmitted to 
$\Delta C(B,T)$.

\section{Conductivity}
Up to this point, we have developed our scaling analysis for 
thermodynamic quantities.  All results have been derived from the 
expression for the free energy density, Eq.~(\ref{eq:f}).  However, the 
description of transport 
measurements, such as the conductivity, requires further 
information.  
Since the equations of motion governing the time 
evolution of the superconductor are not well 
understood, a dynamic scaling ansatz must be postulated.  

Let us 
consider the ohmic conductivity $\sigma$, in order to avoid 
complications arising from current dependence\cite{moloni3}.
For $B=0$, when approaching $T_c$ from above, $\sigma$ 
diverges according to some power law.  Fisher, Fisher, and Huse have 
given arguments leading to the following ansatz\cite{ffh}:
\begin{equation}
\sigma_{\rm fl} \propto t^{-\nu_{xy}(z_{xy}-1)} \quad \quad
\quad B=0 , \label{eq:ansatz}
\end{equation}
where $\sigma_{\rm fl}$ is the fluctuation part of 
$\sigma$, and $z_{xy}$ is the exponent of the supposed 3D~$XY$ 
dynamic universality class.  

Scaling at finite fields then proceeds in the usual way:
\begin{equation}
\sigma =S_b(B,T)+S_k|t|^{-\nu_{xy}(z_{xy}-1)}
\Sigma_{\pm} \left( \frac{B|t|^{-2\nu_{xy}}}{H_k} \right)
 .\label{eq:sigma}
\end{equation}
The first term, 
$S_b(B,T)$, represents the smooth background conductivity.  The 
second term is the fluctuation contribution, where the sample-dependent 
parameter $S_k$ has dimensions of conductivity.  
$\Sigma_+(x)$  [$\Sigma_-(x)$] 
should be universal scaling functions, 
corresponding to $t>0$ [$t<0$].  The background
conductivity term plays a relatively small role 
near the transition line, $T=T_m(B)$, due to the divergence of 
$\sigma_{\rm fl}$.

The functions $\Sigma_\pm (x)$ are not known theoretically, but we 
may deduce their asymptotic behavior.  For $B>0$ and
$T\rightarrow T_c$, the conductivity should be finite, smooth, and 
independent of $t$, leading to
\begin{equation}
\lim_{x\rightarrow \infty} \Sigma_\pm (x)=s_0x^{(1-z_{xy})/2} ,
\label{eq:limitS}
\end{equation}
where $s_0$ is a universal number.  Thus, at $T=T_c$,
\begin{equation}
\sigma_{\rm fl} = (S_ks_0)(B/H_k)^{(1-z_{xy})/2} .
\label{eq:sTc}
\end{equation}
A similar limit can also be taken
for the conjugate fields; we show only the 
results for $H_{\rm ex}$:
\begin{eqnarray}
\sigma_{\rm fl} & = & (S_ks_0)
(H_k\tilde{\Omega} )^{(z_{xy}-1)/2}
\nonumber \\ & & \mbox{}
\times \left[ H_{\rm ex}+2\tilde{H}_0-2\tilde{H}_0^{1/2}
\sqrt{\tilde{H}_0+H_{\rm ex}} \, \right]^{(1-z_{xy})/2} . 
\label{eq:sTcHex}
\end{eqnarray}

The asymptotic behavior of $\Sigma_-(x)$ as 
$T\rightarrow T_m(B>0)$ is of particular interest when 
the transition is continuous.  The multicritical 
description\cite{riedel} involves a crossover from $XY$ 
to glass fluctuations\cite{salamon,yeh,moloni,moloni5,ffh}. 
Although glass fluctuations dominate near $T_m(B>0)$, the XY 
scaling fomula~(\ref{eq:sigma}), is still appropriate.  
Approaching the transition, we find
\begin{equation}
\lim_{x\rightarrow x_{m}} \Sigma_- (x)\propto
\left\{ \begin{array}{ccl}
(x-x_m)^{-\omega} & \quad & x>x_m \\ 
\infty & \quad & x< x_m \\ 
\end{array} \right. ,
\label{eq:limitTm}
\end{equation}
where $x_m$ is the same universal constant as in Section~IV.
The exponent $\omega$ is related to glass, not $XY$ fluctuations.  In 
Ref.~\cite{ffh} it has been argued that
$\omega =\nu_g(z_g-1)$, in analogy 
with Eq.~(\ref{eq:ansatz}), where
$\nu_g$ and $z_g$ are glass exponents.  

\section{Conclusions}
In this paper, we have presented the basic scaling theory of the 
intermediate 
3D~$XY$ transition.  The theory is quite general; it involves only the 
assumption that $B\sim (\mbox{Length})^{-2}$, which is deduced from minimal 
coupling.  We stress that the 
theory should apply to all strongly type-II 
superconductors, including both high-$T_c$ and 
low-$T_c$ varieties.  In particular, we have considered 
the case that the finite field 
transition $T_m(B)$ is continuous, although a similar analysis can be
applied in the case of a first order melting transition\cite{twophase}.

The scaling results can be summarized by noting that the 
$H$-$T$ or $H_{\rm ex}$-$T$ superconducting phase 
diagrams involve two 
regimes, as shown in Fig.~1.  In the outer region, scaling 
is the same for each of the fields $B$, $H$, and $H_{\rm ex}$, up to 
very small correction terms.  To obtain results for the different fields, 
we consider the equations involving only $B$ [for example, (\ref{eq:f}), 
(\ref{eq:BT}), (\ref{eq:Mscale}), (\ref{eq:C}), and (\ref{eq:sigma})], 
then apply the following substitutions:
\begin{equation}
B\leftrightarrow \frac{H}{\Omega_T} \leftrightarrow 
\frac{H_{\rm ex}}{\tilde{\Omega}_T} .\label{eq:outer}
\end{equation}

In the inner region, scaling behaviors differ for $B$, $H$, and 
$H_{\rm ex}$.  This is a nontrivial 
consequence of the conjugate nature of $B$ and $H$, in the 
thermodynamic sense.  Observation of the inner 
region should therefore be regarded as a more stringent test of 
3D~$XY$ scaling.  However, care must be taken to distinguish inner
scaling behavior from inverted $XY$ 
behavior.  The dimensions of the inner region, in the $H$-$T$ 
plane, are given by the characteristic 
field and temperature scales, $H_0$ and $t_1T_c$.  In the $H_{\rm 
ex}$-$T$ 
plane, these become $\tilde{H}_0$ and 
$\tilde{t}_1T_c$.

The high-$T_c$ oxide materials are natural candidates for 
observing both the inner and outer 3D~$XY$ scaling behaviors, due to 
their strongly type-II character, and the prevalence of vortex 
fluctuations near the superconducting transition. However, recent 
experiments demonstrate behavior 
consistent only with the outer region.  
To understand this, it is helpful to obtain estimates for $\tilde{H}_0$, 
$\tilde{H}_{\rm eff}$, 
and $\tilde{t}_1$.   We can make use of two experimental 
analyses.  The first (I), by Hubbard {\it et al.}\cite{hubbard}, involves a 
series of YBa$_2$Cu$_3$O$_{7-\delta}$ single crystals 
with varying $\delta$.  The second (II), by Moloni {\it et 
al.}\cite{moloni}, uses a similar series of thin films.  Both cases 
involve samples with $T_c\simeq 77$~K.  Using the appropriate
demagnetizing factors ($n\simeq 0.84$ for sample~I, and 
$n \simeq 0.992$ for sample~II), 
we can make our estimates.  We will refer to the 
geometry-independent properties of each sample (without tildes) as
``intrinsic." These intrinsic quantities are assumed to be the same for 
both of the 77~K samples.

In Hubbard {\it et al.,} the fluctuation magnetization was plotted
for $T\simeq T_c$, as a function of  the 
field $H_{\rm ex}$.  We can then use Eq.~(\ref{eq:Mfl}) to find 
$\tilde{H}_0$, with the assumption 
$\tilde{\Omega}_c \simeq 1$.  This gives
$\tilde{H}_0\simeq 5\times10^{-7}$~Oe for sample~I, corresponding 
to the intrinsic result 
$H_0\simeq 2\times 10^{-5}$~Oe.   The estimate for sample~II 
then becomes $\tilde{H}_0\simeq 1\times 10^{-9}$~Oe.
 
The relative temperature scale
$\tilde{t}_1$ can be estimated by assuming that along 
the superconducting transition line,
$H_{\rm ex}\simeq \tilde{H}_0$ when $|t|=\tilde{t}_1$.  From 
Eq.~(\ref{eq:HexTp}), this gives 
$\tilde{t}_1 \simeq (\tilde{H}_0/2 H^*)^{1/2\nu_{xy}}$, where
$H^*\equiv \tilde{\Omega}_mx_mH_k$ is another characteristic field of 
experimental significance.  
For sample~II, it was found that
$H^*\simeq 19$~T.  We can then estimate
$\tilde{t}_1 \simeq 1\times 10^{-9}$ for sample~I and 
$\tilde{t}_1 \simeq 1\times 10^{-11}$ for sample~II, with the 
corresponding intrinsic result, 
$t_1 \simeq 2\times 10^{-8}$.  Note that the width of the inner 
scaling region of sample~I becomes 
$\tilde{t}_1T_c\simeq 9\times 10^{-8}$~K.  

The estimates given above for $H_0$ and $t_1$ pertain to
underdoped high-$T_c$ cuprates.  For the case of 
optimally doped cuprates, estimates for $H_0$ and $t_1$
should be somewhat smaller.  Estimates for low-$T_c$
superconductors should be even smaller.
In each case, the smallness of the inner scaling 
region reflects the weak diagmagnetic response of fluctuations 
above $T_m$.  Thus the inner region is probably experimentally 
inaccessible in many cases, due to sample inhomogeneities.  However, 
strong anisotropy effects may improve the situation.  As discussed in 
Sec.~II, an assumed relation ($f_k/\gamma T_c)^2\propto H_k^3$ leads 
to 
$H_0\propto \gamma^2T_c^2$, showing that $H_0$ (and $t_1$) may be 
greatly enhanced in very anisotropic samples.  

It is also possible to estimate of the temperature width, $t_{\rm inv}$, of 
the inverted 3D~$XY$ scaling region.  This may be compared to the size  
of the inner scaling 
region, $t_1$.  Although neither the intermediate nor inverted 
3D~$XY$ models has been solved exactly, the 
crossover may be estimated using the intermediate scaling theory:  it is 
the point where our 
assumption of a uniform magnetic field breaks down.  (See Sec.~II.) 
Approaching the transition, the zero-field screening 
length\cite{kamal,ffh},
$\lambda =\lambda_0|t|^{-\nu_{xy}/2}$, diverges more slowly than the 
3D~$XY$ critical correlation length, $\xi=\xi_0|t|^{-\nu_{xy}}$.  We 
expect a crossover to occur when 
$\kappa = \lambda /\xi \simeq 1/\sqrt{2}$.  The length scale $\xi_0$ of 
3D~$XY$ 
fluctuations is not well known experimentally; 
however, a rough estimate can be obtained by 
using the bare Ginzburg-Landau parameter, which is easily obtained for 
samples such as I and II:  $\kappa_0=\lambda_0/\xi_0\simeq 100$-250.  
We finally obtain
$t_{\rm inv}\simeq2$-$40\times 10^{-8}$.  In this 
rough estimate, the inner and inverted scaling regions are of comparable 
size, and must both be considered in order to correctly interpret fluctuations 
effects very near the 3D~$XY$ critical point.

In contrast to the elusive inner scaling region, the outer region, or 
behavior consistent with it, is readily observed over a
wide temperature range:  $|t|\lesssim 0.5$\cite{moloni}, which is on 
the order of $10^9$-$10^{11}$ times $\tilde{t}_1$.  The difference 
between the two $XY$ temperature 
scales is striking.  However, we note that there is no reason why 
they should be related.  It remains an outstanding theoretical 
problem to provide estimates for the relevant temperature and field 
scales, from 
microscopic considerations.

\section*{Acknowledgments}
We thank K.~Moloni for assistance with the estimates of Sec.~IX.
We also thank M.~Hubbard, S.~Khlebnikov, M.~Salamon, and 
Z.~Te\v{s}anovi\'{c} 
for helpful discussions.
This work was supported by the 
Director for Energy Research, Office of Basic Energy Sciences through 
the Midwest Superconductivity Consortium (MISCON) DOE grant 
\# DE-FG02-90ER45427.

\end{multicols}

\begin{references}
\bibitem{safar}H. Safar, P. L. Gammel, D. A. Huse, D. J. Bishop,
        J. P. Rice and D. M. Ginsberg, \prl 69 (1992) 824;
        E. Zeldov, D. Majer, M. Konczykowski,
        V. B. Geshkenbein, V. M. Vinokur and H. Shtrikman,
        Nature { 375} (1995) 373.
\bibitem{liang}R. Liang, D. A. Bonn and W. N. Hardy, \prl { 76}
        (1996) 835.
\bibitem{roulin}M. Roulin, A. Junod and E. Walker, Science { 273}
        (1996) 1210.
\bibitem{koch}R. H. Koch, V. Foglietti, W. J. Gallagher, G. Koren,
        A. Gupta and M. P. A. Fisher, \prl { 63} (1989) 1511;
        P. L. Gammel, L. F. Schneemeyer and
        D. J. Bishop, \prl { 66} (1991) 953;
        for a more extensive bibliography, see
        G. Blatter, M. V. Feigel'man, V. B. Geshkenbein, A. I. Larkin
        and V. M. Vinokur, \rmp { 66} (1994) 1125.
\bibitem{salamon}M. B. Salamon, J. Shi, N. Overend and M. A.
        Howson, \prb { 47} (1993) 5520;
        M. B. Salamon, W. Lee, K. Ghiron, J. Shi, N. Overend and
        M. A. Howson, Physica A { 200} (1993) 365.
\bibitem{yeh}
        M. A. Howson, N. Overend, I. D. Lawrie and M. B. Salamon,
        \prb { 51} (1995) 11984.
\bibitem{moloni}K. Moloni, M. Friesen, S. Li, V. Souw, P. Metcalf,
        L. Hou and M. McElfresh, \prl { 78} (1997) 3173.
\bibitem{moloni5}K. Moloni, M. Friesen, S. Li, V. Souw, P. Metcalf,
        L. Hou and M. McElfresh, \prb { 56} (1997) 14784.
        M. Friesen, K. Moloni, S. Li, V. Souw, P. Metcalf,
        L. Hou and M. McElfresh, unpublished.
\bibitem{dfisher}It is not necessary that
        $T_c=\lim_{B\rightarrow 0}T_m(B)$, as evidenced in the case 
	of two dimensional superconductors; see D. S. Fisher, 
	\prb { 22} (1980) 190.
\bibitem{inderhees}S. E. Inderhees, M. B. Salamon, J. P. Rice
        and D. M. Ginsberg, \prl { 66} (1991) 232;
        G. Mozurkewich, M. B. Salamon and S. E. Inderhees,
        \prb { 46} (1992) 11914;
        N. Overend, M. A. Howson and I. D. Lawrie,
        \prl { 72} (1994) 3238;
        E. Janod, A. Junod, K.-Q. Wang, G. Triscone, R. Calemczuk
        and J.-Y. Henry, Physica C { 234} (1994) 269;
        M. A. Howson, I. D. Lawrie and N. Overend,
        \prl { 74} (1995) 1888;
        M. Roulin, A. Junod and J. Muller, \prl { 75} (1995) 1869;
        N. Overend, M. A. Howson and I. D. Lawrie,
        \prl { 75} (1995) 1870;
        N. Overend, M. A. Howson, S. Abell and J. Hodby,
        Journ. of Superconductivity { 8} (1995) 677;
        N. Overend, M. A. Howson, I. D. Lawrie, S. Abell, P. J. Hirst,
        C. Changkang, S. Chowdhury, J. W. Hodby, S. E. Inderhees and
        M. B. Salamon, \prb { 54} (1996) 9499.
\bibitem{roulin2}M. Roulin, A. Junod and E. Walker, Physica C
        { 260} (1996) 257.
\bibitem{hubbard}M. A. Hubbard, M. B. Salamon and B. W. Veal,
        Physica C { 259} (1996) 309.
\bibitem{cooper}J.~R.~Cooper, J.~W.~Loram, J.~D.~Johnson, J.~W.~Hodby
	and C.~Changkang, \prl 79 (1997) 1730.
\bibitem{kamal}S. Kamal, D. A. Bonn, N. Goldenfeld, P. J. Hirschfeld,
        R. Liang and W. N. Hardy, \prl { 73} (1994) 1845;
        S. M. Anlage, J. Mao, J. C. Booth, D. H. Wu and J. L. Peng,
        \prb { 53} (1996) 2792;
        Y. Jaccard, T. Schneider, J.-P. Locquet, E. J. Williams,
        P. Martinoli and \O. Fischer, Europhys. Lett. { 34}
        (1996) 281.
\bibitem{anlage}J. C. Booth, D. H. Wu, S. B. Qadri, E. F. Skelton,
        M. S. Osofsky, A. Piqu\'{e}  and S. M. Anlage, \prl { 77}
        (1996) 4438.
\bibitem{ariosa}T. Schneider and D. Ariosa, Z. Phys. B { 89}
        (1992) 267.
\bibitem{kim}J.-T.~Kim, N.~Goldenfeld, J.~Giapintzakis and D.~M.~Ginsberg,
	\prb { 56} (1997) 118.
\bibitem{dasgupta}C. Dasgupta and B. I. Halperin, \prl { 47}
        (1981) 1556;
        M. Kiometzis, H. Kleinert and A. M. J. Schakel, \prl
        { 73} (1994) 1975;
        I. F. Herbut and Z. Te\v{s}anovi\'{c}, \prl { 76}
        (1996) 4588.
\bibitem{ffh}D. S. Fisher, M. P. A. Fisher and D. A. Huse,
        \prb { 43} (1991) 130.
\bibitem{fisher}M. P. A. Fisher, \prl { 62} (1989) 1415;
        D. R. Nelson and V. M. Vinokur, \prl { 68}
        (1992) 2398; \prb { 48} (1993) 13060;
        T. Giamarchi and P. Le Doussal, \prl { 72} (1994) 1530;
        \prb { 52} (1995) 1242.
\bibitem{notezlatko}In particular, we do not consider any scaling details
	of the inverted 3D~$XY$ model class.  We also do not explicitly 
	consider the
	$\Phi$ ($B>0$) transition of Te\v{s}anovi\'{c}, which also belongs
	to the inverted 3D~$XY$ model class\cite{zlatko}; 
	however, within the intermediate 3D~$XY$ critical region, 
	the scaling theory
	of the $\Phi$ transition is entirely consistent with the 
	theory presented here.
\bibitem{twophase}M.~Friesen and P.~Muzikar, cond-mat/9712214, unpublished.
\bibitem{lobb}C. J. Lobb, \prb { 36} (1987) 3930;
        I. D. Lawrie, \prb { 50} (1994) 9456.
\bibitem{zlatko}Z. Te\v{s}anovi\'{c}, \prb { 51} (1995) 16204;
	Z.~Te\v{s}anovi\'{c}, unpublished.
\bibitem{schneider}T. Schneider, Z. Phys. B { 88} (1992) 249;
        Physica B { 222} (1996) 374;
        T. Schneider and H. Keller, \prl { 69} (1992) 3374;
        Physica C { 207} (1993) 366;
        International Jour. of Mod. Phys. B { 8} (1993) 487;
        \prl { 72} (1994) 1133.
\bibitem{lawrie}I. D. Lawrie, \prl { 79} (1997) 131.
\bibitem{numerics}Y.-H.~Li and S.~Teitel, \prb { 47} (1993) 359;
	Y.-H.~Li and S.~Teitel, \prb { 49} (1994) 4136;
	T.~Chen and S.~Teitel, \prl { 74} (1995) 2792;
	A.~K.~Nguyen, A.~Sudb\o and R.~E.~Hetzel, \prl { 77} (1996) 1592;
	T.~Chen and S.~Teitel, \prb { 55} (1997) 11766;
	T.~Chen and S.~Teitel, \prb { 55} (1997) 15197;
	S.~Ryu and D.~Stroud, \prl { 78} (1997) 4629;
	A.~E.~Koshelev, \prb { 56} (1997) 11201;
	X.~Hu, S.~Miyashita and M.~Tachiki, \prl { 79} (1997) 3498;
	A.~K.~Nguyen and A.~Sudb\o, \prb { 57} (1998) xxx;
	S.~Ryu and D.~Stroud, cond-mat/9712246, unpublished;
	A.~K.~Nguyen and A.~Sudb\o, cond-mat/9712264, unpublished.
\bibitem{goldenfeld}For a discussion, see N. Goldenfeld,
        { Lectures on Phase Transitions and the Renormalization
        Group} (Addison-Wesley, Reading, MA, 1992).
\bibitem{note1}We have used the standard definition of specific heat
        fluctuations:  $C=-T\, \partial^2f/\partial T^2 \propto t^{-\alpha}$.
	Based upon the dimensional statements $B\sim (\mbox{Length})^{-2}$ and
	$t\sim (\mbox{Length})^{-1/\nu}$, Eq.~(\ref{eq:ftilde}) can be
	derived according to Ref.~\cite{privman}.
\bibitem{privman}See V. Privman, P. C. Hohenberg and A. Aharony,
        in: { Phase Transitions and Critical Phenomena 14},
        C. Domb and J. L. Lebowitz, ed. (Academic, London, 1991)
        p.~1, and references therein.
\bibitem{blatter}R. A. Klemm and J. R. Clem, \prb { 21} (1980) 1868;
        G. Blatter, V. B. Geshkenbein and A. I. Larkin,
        \prl { 68} (1992) 875.
\bibitem{riedel}E. K. Riedel, \prl { 28} (1972) 675;
        for a review, see I. D. Lawrie and S. Sarbach, in:
        { Phase Transitions and Critical Phenomena 9,} 
        C. Domb and J. L. Lebowitz, ed. (Academic, London, 1984)
        p.~1.
\bibitem{note4}Nonellipsoidal geometry can have important
        thermodynamic consequences
        for the phase transition; for example, see,
        D. E. Farrell, E. Johnston-Halperin, L. Klein, P. Fournier,
        A. Kapitulnik, E. M. Forgan, A. I. M. Rae, T. W. Li,
        M. L. Trawick, R. Sasik and J. C. Garland, \prb { 53}
        (1996) 11807.
\bibitem{landau}L. D. Landau and E. M. Lifshitz, { Electrodynamics
        of Continuous Media} (Pergamon, Oxford, 1960).
\bibitem{degennes}P. G. de Gennes, { Superconductivity
        of Metals and Alloys} (Addison-Wesley, Redwood City, CA,
        1989).
\bibitem{note5}An analysis along these lines has been performed in
        Ref.~\cite{hubbard}.  Unfortunately, a check of the relation
        $f_kH_k^{-3/2}\propto \gamma T_c$ is impossible, due
	to large error bars.
\bibitem{zinn}Theory:  J. C. LeGuillou and
        J. Zinn-Justin, J. Phys. (Paris) { 46} (1985) L137;
        experiments (in $^4$He): J. A. Lipa and T. C. P. Chui, \prl
        { 51} (1983) 2291.
\bibitem{moloni3}Aspects of the scaling of nonlinear conductivity are
        considered in Ref.~\cite{moloni5}.
\end{references}
\end{document}